\documentclass[aps,prd,11pt,notitlepage,nofootinbib,superscriptaddress,showkeys,showpacs]{revtex4-1}
\usepackage{amsmath,amssymb,amsthm,latexsym}
\usepackage{stmaryrd,wasysym,upgreek,mathrsfs,dsfont}
\usepackage[english]{babel}
\usepackage{graphicx,color}
\usepackage{xspace}
\usepackage{graphicx}
\usepackage{pifont,dsfont}
\usepackage{marvosym}
\usepackage{slashed}
\usepackage{subfigure}

\newcommand{\be}{\begin{equation}}
\newcommand{\ee}{\end{equation}}
\newcommand{\bqa}{\begin{eqnarray}}
\newcommand{\eqa}{\end{eqnarray}}
\newcommand{\bea}{\begin{eqnarray}}
\newcommand{\eea}{\end{eqnarray}}

\newcommand{\Tr}{{\rm Tr}}

\newtheorem{proposition}{Proposition}

\newcommand{\cF}{{\cal F}}

\newcommand{\cG}{{\cal G}}

\begin{document}

\title{\Large \bf The Ising Model on Random Lattices in Arbitrary Dimensions}

\author{{\bf Valentin Bonzom}}\email{vbonzom@perimeterinstitute.ca}
\author{{\bf Razvan Gurau}}\email{rgurau@perimeterinstitute.ca}
\affiliation{Perimeter Institute for Theoretical Physics, 31 Caroline St. N, ON N2L 2Y5, Waterloo, Canada}
\author{{\bf Vincent Rivasseau}}  \email{vincent.rivasseau@gmail.com}
\affiliation{Laboratoire de Physique Th\'eorique, CNRS UMR 8627, Universit\'e Paris XI,  F-91405 Orsay Cedex, France}

\date{\small\today}

\begin{abstract}
\noindent We study analytically the Ising model coupled to random lattices in dimension three and higher. 
The family of random lattices we use is generated by the large $N$ limit of a colored tensor model 
generalizing the two-matrix model for Ising spins on random surfaces. We show that, in the continuum limit, the spin system does not exhibit a phase transition at finite temperature, in agreement with numerical investigations. Furthermore we outline a general method to study critical behavior in colored tensor models.

\end{abstract}

\medskip

\noindent  Pacs numbers: 02.10.Ox, 04.60.Gw, 05.40-a

\maketitle

\section{Introduction}

Random matrix models  \cite{Kazakov:1985ds,mm,Di Francesco:1993nw} provide a statistical theory of  random  discretized Riemann surfaces. The amplitudes of the ribbon Feynman graphs of their perturbative expansion support a $1/N$ expansion 
\cite{'tHooft:1973jz} (where $N$ is the size of the matrices) indexed by the genus of the surfaces.
In the large $N$ limit the planar graphs corresponding to surfaces of spherical topology dominate. 
The genus is related (by the Gauss-Bonnet theorem) to the Einstein-Hilbert action on the two-dimensional surfaces 
and the large $N$ parameter appears as the inverse Newton constant.

The planar graphs which lead this expansion proliferate exponentially, like $K^n$, with the number of vertices $n$. 
The sum over random lattices of spherical topology is thus convergent for small enough coupling constant 
and the planar free energy is finite. Planar graphs can be counted precisely through algebraic equations \cite{Brezin:1977sv}, as they are related to trees \cite{Tut,Sch,BMS}, the universal structures behind such equations. When the coupling constant grows, the free energy becomes dominated by graphs with a large number of vertices and exhibits a critical behavior. It is in this regime that the system reaches its continuum limit and the critical exponents can be evaluated.

In the seminal paper \cite{Kazakov:1986hu}, followed by \cite{Boulatov:1986sb, Brezin:1989db}, Kazakov et al.
solved the two-dimensional Ising model on random geometries using a two matrix model. 
Surprisingly this solution turned out to be simpler than the one on a fixed lattice. The importance of this work was further
enhanced by the discovery of the KPZ correspondence \cite{Knizhnik:1988ak, david2, DK} which relates two-dimensional conformal field
theories coupled or not to Liouville gravity, and by the introduction of the double scaling limit of random matrices which combines all genera \cite{double,double1,double2}. 
The back and forth navigation between two dimensional statistical systems on fixed and random geometries led to early computations of novel critical 
exponents  \cite{Dup}. Their values have been later confirmed by rigorous probabilists, through many developments in particular rewarded 
by the Fields medals of W. Werner and S. Smirnov.
The transformation of the two dimensional statistical mechanics landscape was so deep that it could be considered a change of paradigm. Statistical 
models on random geometry appear somehow more fundamental and ordinary statistical physics on a fixed lattice as a quenched version in which
the fluctuating geometry has been frozen.

It would be highly desirable to generalize such ideas and results to more than two dimensions. Random matrices generalize in higher dimensions 
to random tensors \cite{ambj3dqg,sasa1,mmgravity}, whose perturbative expansion performs a sum over random higher dimensional geometries. But 
until recently the key to analytical rather than numerical results, namely the $1/N$ expansion, was missing.

That situation has changed with the discovery of such a $1/N$ expansion \cite{Gur3,GurRiv,Gur4} for {\it colored} 
\cite{color,PolyColor,lost} random tensors. The amplitude of their graphs supports a $1/N$ expansion indexed by
the {\it degree}, a positive integer, which plays in higher dimensions the role the genus played in two dimensions. 
The leading order graphs, baptized {\it melonic} \cite{Bonzom:2011zz}, triangulate the $d$-dimensional sphere. 
They form a summable series and map to colored d-ary trees \cite{Bonzom:2011zz,Gurau:2011tj}. When the coupling constant approaches its critical 
value, the free energy exhibits a critical behavior and,  like in matrix models, the colored tensor models reach their continuum limit
dominated by triangulations with an infinite number of simplices.  The entropy exponent of the melonic series, analogous to the string 
susceptibility $\gamma_{\rm string} = -1/2$ of the 1-matrix model for the pure gravity universal class, is 
$\gamma_{\rm melons} = 1/2$, \cite{Bonzom:2011zz}.

Colored random tensors are a promising tractable discretization of quantum gravity in three and more dimensions and the subject is 
developing fast \cite{sefu2,Baratin:2011tg,Geloun:2011cy,Bonzom:2011br}.
The understanding of the leading (melonic) order of colored tensor models allows the study of the coupling of statistical systems 
to random geometries in arbitrary
dimension. Concretely one needs to develop an algorithm for solving tensor models and study the critical behavior
of various particular examples. 
As some of the more powerful tools of matrix models, like the reduction to eigenvalues, are absent in tensor models, 
the algorithm we present here relies on combining Schwinger-Dyson (SD) equations with a factorization property 
characteristic of melonic graphs
\cite{Bonzom:2011zz}. 

We illustrate our method on the Ising model coupled to random melonic triangulations. It is generated by a (colored) two-tensor model, 
the straightforward generalization of the matrix model studied by Kazakov \cite{Kazakov:1986hu} in arbitrary dimensions. It is presented in the Section \ref{sec:color}. Unlike in two dimensions, we find in the Section \ref{sec:analysis} that the model does 
not exhibit a phase transition in the continuum limit. The large $N$ limit corresponds to the limit of zero bare Newton's constant. As such it is only a limiting case of numerical simulations. However, our results are in agreement with numerical simulations for small Newton's constant \cite{Ambjorn:1992iz} associated to the (cold) branched-polymer phase\footnote{The numerical studies
consider triangulations of the sphere which are not necessarily colorable or melonic. This is of no concern as many such ensembles are 
expected to belong to the same universality class. As already shown in \cite{Bonzom:2011zz}, the 
melonic family shares a number of features with the branched polymers.}. An open question is therefore to understand why and how the progressive
freezing of the random geometry would let the ordinary Ising phase transition appear.

\section{Colored Random Tensor Models} \label{sec:color}

\subsection{Ising Model on Random Surfaces as a Two-Matrix Model}

A single random matrix model in the large $N$ limit can be used to describe the behavior of pure two dimensional Liouville gravity,
with string susceptibility (or entropy) exponent $\gamma = -1/2$ \cite{Kazakov:1985ds,mm,Di Francesco:1993nw}. The free energy $F$ expands 
in the coupling constant $g$ as $F \simeq \sum_n   F_n   g^n$, where $F_n$ counts the number of planar lattices with $n$ vertices (of given 
valence). The entropy exponent $\gamma$ is deduced from the asymptotic of $F_n$: if $F_n \simeq n^{-b}  K^n$,  then,
for $g$ close to the critical coupling $g_c$, $F\simeq (g_c-g)^{2-\gamma}$ with $\gamma = -b+3$. The continuum, large volume, 
limit dominated by lattices with a infinite number of vertices is reached when tuning $g$ to $g_c$. 

In \cite{Kazakov:1986hu} the critical behavior of the two dimensional Ising model on random planar quadrangulations is analyzed 
through a model of two coupled random hermitian matrices $X$ and $Y$ representing respectively 
the up and down Ising spins. The free energy of the model is
\be
F(c,g) = N^{-2}    \log \int dX dY \exp \Tr - \bigl[U^2 + Y^2 -2c XY  +\frac{g}{N} X^4  + \frac{g}{N} Y^4   \bigr ]
\ee
where $c= e^{-2\beta J}$, $\beta$ being the inverse temperature, and $J>0$ the Ising ferromagnetic coupling.
The  free energy and coupling constant can be solved explicitly in parametrized form as
\bea
F(c,z)  &= & \frac{z^2}{2g^2(z)} \Bigl[  \frac{z-1}{2(3z-1)^3}  + c^2\frac{z+1}{3z-1}  + \frac{c^4}{2  (3z^4 - 3z^2 + 1) } \Bigr] \crcr
 &&- \frac{z}{g(z)} \Bigl[ \frac{1}{3z-1} + c^2 (1-z^2) \Bigr] 
 + \frac{1}{2}\ln \frac{z(1-c^2)}{g(z)} + \frac{3}{4}\;, \crcr
g(c,z) &=& \frac{z}{(1-3z)^2} - c^2z + 3c^2 z^3 \; .
\eea

The function $F(c,g)$ becomes critical in $g$ when
\be  
\frac{\partial g}{\partial z} = 0 \;,
\ee
and the random matrix model reaches its continuum limit. The roots of this equation are 
\bea
z_0 = -\frac{1}{3}\;, \qquad z_{1,2}= \frac{1}{3}(1 \mp c^{-1/2})\;, \qquad z_{3,4}= \frac{1}{3}(1 \mp i c^{-1/2})\;.
\eea
In the physical interval of temperature, $0< c <1$, only the two first roots $z_{0,1}$ are relevant (as the others are well separated),
and collapse, $z_0=z_1$, at $c=1/4$. The corresponding critical couplings are
\bea
g_0(c) = -\frac{1}{12}  + \frac{2c^2}{9}, \qquad g_{1}(c)   = -\frac{2\sqrt c}{9}   (\sqrt c -1)^2(\sqrt c +2)\;.
\eea
The first coupling, $g_0(c)$, is the critical line (of the large volume limit) at temperature $c<1/4$.
The second coupling, $g_c(c)$, is the critical line at temperature $c>1/4$. At $c=1/4$, the two 
lines meet and a phase transition between the low and high temperature phases of the continuous Ising model on a random 
lattice occurs. 
Evaluating the free energy in the low and high temperature phases, $F(c,z_0(c))$ and $F(c,z_1(c))$ one finds that first and second 
derivatives of $F$ with respect to $c$ are continuous at the phase transition, but not the third, 
hence the transition is third order.

Note that $F_n$ is the canonical partition function of Ising spins on a random planar quadrangulation with $n$ vertices. Away from the 
critical temperature $c=1/4$, $F_n$ behaves like $n^{-b}K^n$ with $b = 7/2$, so that the corresponding susceptibility exponent 
is $\gamma = -b+3 = -1/2$, 
just like pure two dimensional Liouville gravity. As the two roots $z_0$ and $z_1$ meet at the critical temperature, the second derivative 
$\partial^2 g/\partial z^2_{|c=1/4}$ vanishes and $F_n$ changes its large $n$ behavior to $n^{-b'}K^n$ with $b'=10/3$ and 
corresponding susceptibility exponent $\gamma=-1/3$.

\subsection{The independent identically distributed 1-tensor model}

We denote $\vec k_i$, for $i=0,\dotsc,D$, the $D$-uple of integers
 $\vec k_i = (k_{ii-1},\dotsc, k_{i0},\; k_{iD}, \dotsc, k_{ii+1}) $, with
$k_{ik}=1,\dotsc, N$. This $N$ is the size of the tensors and the large $N$ limit defined in
\cite{Gur3, GurRiv, Gur4} represents the limit of infinite size tensors.
We set $k_{ij} = k_{ji}$. Let $\bar \psi^i_{\vec k_i},\; \psi^i_{\vec k_i}$, with $i=0,\dotsc, D$, be $D+1$ couples of complex
conjugated tensors with $D$ indices. The independent identically distributed (i.i.d.) colored tensor
model in dimension $D$ \cite{color,lost,Gur4} is defined by the partition function
\begin{gather}
\nonumber e^{- N^D F^{\rm i.i.d.}_N(\lambda,\bar\lambda)} = Z^{\rm i.i.d.}_N(\lambda, \bar{\lambda}) = \int \, d\bar \psi \, d \psi
\ e^{- S^{\rm i.i.d.} (\psi,\bar\psi)} \; , \\
S^{\rm i.i.d.} (\psi,\bar\psi) = \sum_{i=0}^{D} \sum_{\vec k} \bar \psi^i_{\vec k_i} \psi^i_{\vec k_i}  +
\frac{\lambda}{ N^{D(D-1)/4} } \sum_{\vec k} \prod_{i=0}^D \psi^i_{ \vec k_i } +
\frac{\bar \lambda}{ N^{D(D-1)/4} } \sum_{\vec k}
\prod_{i=0}^D \bar \psi^i_{ \vec k_i } \; . \label{eq:iid}
\end{gather}
$\sum_{\vec k}$ denotes the sum over all indices $\vec k_i$ from $1$ to $N$.

The partition function of equation \eqref{eq:iid} is evaluated by colored stranded Feynman graphs
\cite{color,lost}. The colors $i$ of the fields $\psi^i, \bar{\psi}^i$ induce important
restrictions on the combinatorics of stranded graphs. Note that we have two types of vertices,
say one of positive (involving $\psi$) and one of negative (involving $\bar \psi$) orientation.
The lines always join a $\psi^i$ to a $\bar{\psi}^i$ and possess a color index.
The tensor indices $k_{ij}$ are preserved along the strands. The amplitude of a graph with $2p$
vertices and $\cF$ faces is \cite{Gur4}
\bea\label{eq:ampli}
A(\cG)= (\lambda\bar\lambda)^p N^{-p \frac{D(D-1)}{2}+ \cF } \; .
\eea

Any Feynman graphs $\cG$ of this model is a simplicial pseudo manifold \cite{lost}. The colored
tensor models provide thus a statistical theory of random triangulations in dimensions $D$, generalizing random matrix models. The vertices,
 edges and faces (closed strands) of the graph represent the $D$, $(D-1)$ and $(D-2)$-simplices.
The simplices of the pseudo-manifold are identified by the $q$-bubbles of the
graph, for $q=0,\dotsc,D$, i.e. the maximally connected subgraphs made of lines with $q$ fixed colors. Bubbles are local objects (in the sense that each of them 
corresponds to a single simplex of the triangulation) and encode the cellular structure of the pseudo manifold.  The $0$-bubbles, $1$-bubbles 
and $2$-bubbles of a graph are its vertices, lines and faces.

The amplitude \eqref{eq:ampli} of a graph computes in terms of its {\it degree} $\omega(\cG)$ \cite{Gur4} as 
\be
A(\cG)= (\lambda \bar \lambda)^p\,N^{D - \frac{2}{(D-1)!}\omega(\cG)}\;,
\ee
where $p$ is half the number of vertices of $\cG$. The degree is a positive integer, hence the expansion is dominated by 
graphs of degree $\omega(\cG)=0$. Such graphs, dual to triangulations of the $D$-sphere \cite{Gur4} in any dimension,
are the analogs of the planar graphs of matrix models in higher dimensions. They are called \emph{melons} 
\cite{Bonzom:2011zz} and we will restrict our analysis to the melonic sector in the rest of this paper. The melonic graphs can be fully 
characterized at the combinatorial level by a factorization property \cite{Bonzom:2011zz} which we will recall and use in the sequel.

The free energy $F_N^{\rm i.i.d.}(\lambda\bar\lambda)$ has been analyzed in details in \cite{Bonzom:2011zz}. It is analytic for small
values of the coupling $g=\lambda\bar\lambda$. When $g$ 
approaches $g_{c} = \frac{D^D}{(D+1)^{D+1}}$, the free energy becomes non-analytic and 
the i.i.d. tensor model reaches its continuum, large volume limit. The singular part of the free energy is
 $F_{\rm sing}\sim (g_c-g)^{\frac{3}{2}}$. These results rely on a self-consistency equation for the connected melonic 2-point function 
$U^{\rm i.i.d.}$ and its relation with the free energy
\be
U^{\rm i.i.d.} = 1+ g \,\bigl( U^{\rm i.i.d}\bigr)^{D+1}\;, \qquad  g\partial_g F^{\rm i.i.d } = 1-U^{\rm i.i.d} \; ,
\ee
which we will generalize below.

\subsection{The random colored two-tensor model}

We turn now to a model with two complex tensors, both colored, 
$(X^i,Y^i)$ (and $(\bar X^i,\bar Y^i)$). We consider the partition function
\begin{eqnarray}
\nonumber e^{ - N^D F(x,\bar x, y, \bar y, c)} &=&
Z(x,\bar x, y, \bar y, c) = \int \, d\bar X \, d X  d\bar Y \, d Y
\ e^{- S (X, \bar X , Y  \bar Y )} \; , \nonumber  \\
S (X, \bar X , Y,  \bar Y ) &=& \sum_{i=0}^{D} \sum_{\vec k}   \biggr[ \bar X^i_{\vec k_i} X^i_{\vec k_i}
+ \bar Y^i_{\vec k_i} Y^i_{\vec k_i}  -
c  \bar X^i_{\vec k_i} Y^i_{\vec k_i}  - c \bar Y^i_{\vec k_i} X^i_{\vec k_i}   \biggr]
\nonumber  \\
&+& \frac{x}{ N^{D(D-1)/4} } \sum_{\vec k} \prod_{i=0}^D X^i_{ \vec k_i } +
\frac{\bar x}{ N^{D(D-1)/4} } \sum_{\vec k}
\prod_{i=0}^D \bar X^i_{ \vec k_i } \; \nonumber  \\
&+& \frac{y}{ N^{D(D-1)/4} } \sum_{\vec k} \prod_{i=0}^D Y^i_{ \vec k_i } +
\frac{\bar y}{ N^{D(D-1)/4} } \sum_{\vec k}
\prod_{i=0}^D \bar Y^i_{ \vec k_i } \; .
 \label{eq:ising-tensor}
\end{eqnarray}
Again $\sum_{\vec k}$ denotes the sum over all indices $\vec k_i$ from $1$ to $N$.

The Feynman graphs of the action \eqref{eq:ising-tensor} are identical with the ones of
the i.i.d. model, up to the fact that a vertex now involves either $X$ or $Y$ tensors (both coming with 
positive and negative orientations). Tensor indices $k_{ij}$ and colors are preserved along the strands, and the propagator between the $X$ and $Y$ sectors is
\be
C = \frac1{1-c^2} \begin{pmatrix} 1&c\\c&1\end{pmatrix} \;.
\ee
Lines in a graph can join a field $\bar X$ (resp. $\bar Y$) to a field $X$ (resp. $Y$), with weight $1/(1-c^2)$,
or a field $\bar X$ (resp. $\bar Y$) to $Y$ (resp. $X$), with weight $c/(1-c^2)$. This is the only coupling 
between $X$-tensors and $Y$-tensors. As in the i.i.d. model, the free energy organizes 
in powers of $1/N$, with melonic graphs dominating the large $N$ limit, so that
\bea
F (x,\bar x, y, \bar y, c)  
=\sum_{\text{melons }\cG} s(\cG)\ x^{p_x} \bar x ^{p_{\bar x}} y^{p_y} \bar y^{p_{\bar y} }
\, \frac{ c^{L_{\bar XY} + L_{X \bar Y}} }{(1-c^2 )^{L} }  + O(N^{-1}) \; ,
\eea
with $p_x$ (resp. $p_{\bar x}$, $p_y$ and $p_{\bar y}$) the number of $X$ (resp. $\bar X$ $Y$ and $\bar Y$) vertices, 
$L$ is the total number of lines, $L_{\bar X Y}$ (resp. $L_{X\bar Y}$) the number of lines from a $\bar X$ to a $Y$
(resp. from a $X$ to a $\bar Y$) and $s(\cG)$ a symmetry factor. We have $L=(D+1)(p_x+p_y)$, $p_x+p_y = p_{\bar x} + p_{\bar y}$.

The $X$-vertices and $Y$-vertices represent the Ising spins on random lattices. One, say $X$, mimics the 
random world of the up spins, while the other the random world of the down spins. On a graph $\cG$, the total number 
of vertices $n$ is the number of Ising spins. Through the propagator, two spins up or down are coupled with the 
same weight $1/(1-c^2)$ and two opposite spins with $c/(1-c^2)$, from which one can extract the Ising coupling $\beta J$ 
where $\beta$ is the inverse temperature. Taking $c\in[0,1]$ corresponds to the ferromagnetic model $J>0$. 
Setting $x=\bar x$ and $y=\bar y$, the free energy of the tensor model equates the grand canonical partition function 
of the Ising model on random lattices,
\begin{align}
\nonumber Z(\mu, \beta J,\beta h) &\,=\, \sum_{n} e^{-n\mu}\ Z_n(\beta J, \beta h)\;,\\
Z_n(\beta J, \beta h)\ &= \sum_{\cG, 2(p_x+p_y) = n} s(\cG)\, \sum_{s_i=\pm 1} e^{\beta J \sum_{<i,j>} s_i s_j +\beta h\sum_{i} s_i }\;.
\end{align}
$Z_n$ is the canonical partition function for the Ising system on random lattices with a fixed number of 
spins $n$, $h$ is the magnetic field and $\mu$ the chemical potential. The relationship between the Ising parameters 
and those of the tensor model is
\be
e^{ - 2 \beta J} =c\;,\qquad e^{-\mu} = \Bigl( \frac{ 1 } { c^{D+1}  (1-c^2)^{ D+1 } }
\sqrt{x y}\Bigr)^{1/2}   \;, \quad e^{\beta h} = \left(\frac{ x}{ y  } \right)^{1/2}
\; .
\ee

\section{Analysis of the system} \label{sec:analysis}

\subsection{Method}

The analysis of the critical behavior of our two tensor model is done in three steps
\begin{itemize}
 \item We derive SD equations relating the derivatives of the free energy to the connected 2-point functions of the tensor model.
 \item We evaluate the 2-point functions using the SD equation relating them to the self-energy $\Sigma$ (1PI amputated 2-point function).
In turn, the free energy $\Sigma$ writes in the melonic sector in terms of the full 2-point functions thanks to the melonic 
factorization at large $N$ (the defining property of melonic graphs \cite{Bonzom:2011zz}).
 \item We send the random lattices to criticality while keeping the temperature $c$ fixed. This is the regime where the 
free energy (the grand canonical partition function of the Ising model) is dominated by the graphs with many vertices, 
hence corresponds to the continuum Ising model. We analyze the behavior of the Ising partition function 
as a function of the temperature $c$.
\end{itemize}

\subsection{Derivatives of the grand-canonical free energy}

As the model depends on two complex tensors, there are four types of 2-point functions,
noted $\big\langle\bar X X \big\rangle$, $\big\langle\bar Y Y\big\rangle$,
$\big\langle\bar X Y\big\rangle$ and $\big\langle\bar Y X \big\rangle$ one must deal with.
They are necessarily connected and, due to the conservation of the indices along the faces of the graph,
have the index structure
\bea\label{eq:2point}
\big\langle \bar X ^i_{\vec{n}_i}\,X^i_{\vec{p}_i}  \big\rangle_{\rm c} =
  \delta_{\vec{n}_i,\vec{p}_i} \,  \big\langle \bar X X \big\rangle\; ,
\eea
where $\delta_{\vec{n}_i,\vec{p}_i} $ denotes $ \prod_{k\neq i} \delta_{n_{ik} p_{ik}}$.
All correlation functions depend on the variables $ x , \bar x, y,\bar y, c$.

\begin{proposition}
 The 2-point functions and the derivative of the free energy are related by
\bea
\frac{1}{2} \Bigl( \bar x \frac{\partial}{\partial \bar x} +  x \frac{\partial}{\partial x}
+\bar y \frac{\partial}{\partial \bar y}+ y \frac{\partial}{\partial y} \Bigr) F
 = 2 -  \big\langle \bar X X \big\rangle -  \big\langle \bar Y Y \big\rangle
  + c \big\langle \bar X Y \big\rangle + c \big\langle \bar Y X \big\rangle\;.
\eea

\end{proposition}
{\bf Proof:} Using trivial identities of the type
\bea
0= \frac{1}{Z}\sum_{\vec k_i} \int  \frac{\delta}{ \delta \bar X_{\vec k_i}}
\Big{(}  c\bar X_{ \vec k_i} e^{-S}\Big{)} \; ,
\eea
we get
\bea
&&0= N^D - N^D  \big\langle \bar X X \big\rangle + c  N^D \big\langle \bar X Y \big\rangle
- \frac{ \bar x}{N^{D(D-1)/4 } } \big\langle \sum_n  \bar X_{\vec k_i}
 \prod_{j\neq i} \bar X^j_{\vec k_j } \big\rangle\;, \crcr
&&0= N^D - N^D  \big\langle \bar X X \big\rangle + c  N^D \big\langle \bar Y X \big\rangle
- \frac{  x}{N^{D(D-1)/4 } } \big\langle \sum_n   X_{\vec k_i}
 \prod_{j\neq i} X^j_{\vec k_j } \big\rangle\;,
\eea
that is
\bea
&&  1 - \big\langle \bar X X \big\rangle + c \big\langle \bar X Y \big\rangle
 = \bar x \;  \frac{\partial F}{\partial \bar x}\;, \crcr
&& 1 - \big\langle \bar X X \big\rangle + c \big\langle \bar Y X \big\rangle
 =  x \;  \frac{ \partial F}{\partial x}\;.
\eea
This implies
\bea
\frac{1}{4} \Bigl( \bar x \frac{\partial}{\partial \bar x} +  x \frac{\partial}{\partial x}
+\bar y \frac{\partial}{\partial \bar y}+ y \frac{\partial}{\partial y} \Bigr) F
 = 1 - \frac{1}{2} \big\langle \bar X X \big\rangle - \frac{1}{2} \big\langle \bar Y Y \big\rangle
 + \frac{1}{2} c \big\langle \bar X Y \big\rangle + \frac{1}{2} c \big\langle \bar Y X \big\rangle\;.
\eea
\qed

Also note that
\bea
c\,\partial_c F = N^{-D}(- c)\partial_c \ln Z =
(- c) \Bigl( \big\langle \bar X Y \big\rangle +  \big\langle \bar Y X \big\rangle   \Bigr)\;.
\eea
If $x = \bar x = y =\bar y = \sqrt{g}$, the free energy becomes a function of only one variable (denoted by abuse 
of notation $F(c,g)$) and equates the partition function of an Ising model in zero magnetic field.
Denoting $ U = \big\langle \bar X X \big\rangle =  \big\langle \bar Y Y \big\rangle$ and
$V = \big\langle \bar X Y \big\rangle =  \big\langle \bar Y X \big\rangle$ we get
\bea
 g\partial_g F(c,g) = 2\bigl(1-U+cV\bigr)\;, \qquad c\,\partial_c F(c,g) =  -2cV\;.
\eea

\subsection{Large N Melonic Factorization}

The second SD equation we are using is the classical one relating the connected 2-point function and
 the self-energy $\Sigma$. Taking into account that the propagators connecting two self-energy insertions $\Sigma$
correspond either to $X$ or to $Y$, we can write it in matrix form as
\bea \label{eq:geom}
&& \begin{pmatrix}
     \langle \bar X X \rangle & \langle \bar X Y \rangle \\
     \langle \bar Y X \rangle & \langle \bar Y Y \rangle
 \end{pmatrix} = C +C\Sigma C +C\Sigma C \Sigma C + \dotsb = C \frac{1}{1-\Sigma C}\;, \crcr
&&  \text{for}\qquad
    \Sigma=\begin{pmatrix}
            \langle X \bar X  \rangle_{1PI} &   \langle X \bar Y \rangle_{1PI} \\
            \langle  Y \bar X \rangle_{1PI} & \langle Y \bar Y  \rangle_{1PI}
           \end{pmatrix} \; ,
\eea
where all 2-point functions carry the same color index.

\begin{figure}[htb]
\begin{center}
 \includegraphics[width=6cm]{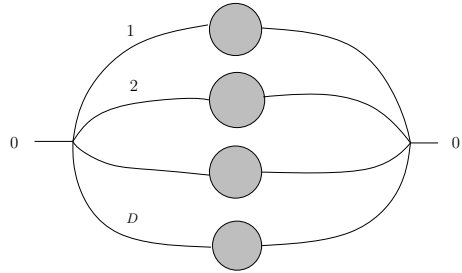}
\caption{Definition of melonic 1PI graphs. The self-energy $\Sigma$ factors into connected 2-point functions 
corresponding to the internal lines.}
\label{fig:1PImelons}
\end{center}
\end{figure}

The simplest melonic 2-point graph \cite{Bonzom:2011zz} is the graph with two vertices connected by 
$D$ lines. Intuitively, more complicated melons are obtained by inserting repeatedly the same motive 
(i.e. two vertices connected by $D$ lines) on lines. In practice this means that the self-energy $\Sigma$ 
in the melonic sector factors as the convolution of $D$ connected 2-point functions, one for each color. 
This is the defining property of melons \cite{Bonzom:2011zz} and is represented in Figure \ref{fig:1PImelons}.
The key consequence of the melonic factorization is that it makes it possible to close the Equation \eqref{eq:geom} 
for the full 2-point function. Indeed,
\bea
 \Sigma  =\begin{pmatrix}
            x\bar x \langle X \bar X  \rangle^D &  x \bar y \langle X \bar Y \rangle^D \\
            y \bar x \langle  Y \bar X \rangle^D & y \bar y  \langle Y \bar Y  \rangle^D
           \end{pmatrix} \; ,
\eea
which can be plugged in \eqref{eq:geom} to yield
\bea
 \begin{pmatrix}
     \langle \bar X X \rangle & \langle \bar X Y \rangle \\
     \langle \bar Y X \rangle & \langle \bar Y Y \rangle
 \end{pmatrix} \Bigl[ 1 -
     \begin{pmatrix}
            x\bar x \langle X \bar X  \rangle^D &  x \bar y \langle X \bar Y \rangle^D \\
            y \bar x \langle  Y \bar X \rangle^D & y \bar y  \langle Y \bar Y  \rangle^D
           \end{pmatrix}
       \frac{1}{1-c^2 }\begin{pmatrix}
          1 & c \\
           c & 1 \\
       \end{pmatrix}
 \Bigr] =
    \frac{1}{1-c^2 }\begin{pmatrix}
          1 & c \\
           c & 1 \\
       \end{pmatrix} \; .
\eea
At zero magnetic field, the matrix equation reduces to two independent equations
\bea\label{eq:syst}
&& U - cV - g \Bigl( U^{D+1} + V^{D+1} \Bigr) =1 \crcr
&& V - cU  - g UV \Bigl( U^{D-1} + V^{D-1}\Bigr) =0 \; .
\eea

\subsection{Critical behavior}

The solution for the noninteracting model (g=0) at arbitrary temperature is
\bea\label{eq:condinit}
   U(c,0) = \frac{1}{1-c^2} \qquad V(c,0) = \frac{c}{1-c^2} \; ,
\eea
which is a boundary condition for the system of equations \eqref{eq:syst}.
Multiplying the first equation by $U$, the second one by $V$ and subtracting
we get
\bea\label{eq:VdeU}
 V =  \sqrt{ \frac{U^2 -U - gU^{D+2}}{ 1-gU^D } } = U \sqrt{1- \frac{1}{U (1-gU^D)}} \; .
\eea
where we chose the determination of the square root verifying the boundary condition. Indeed,
 $U= \langle \bar X X \rangle$ is positive, while $V(c,0)>0$ for all $c>0$, and $V$ vanishes only if
 either $U=0$, or $c=0$. The square-root in equation \eqref{eq:VdeU}, equals $V/U$ and is smaller than one, since
 $2(U-V)= \langle (\bar X - \bar Y) ( X - Y) \rangle \geq 0$. Substituting back into \eqref{eq:syst} we obtain 
\bea\label{eq:fund}
 c(U,g) = (1-gU^D) \sqrt{1- \frac{1}{U (1-gU^D)} } - gU^D \Big{(} 1- \frac{1}{U (1-gU^D)} \Big{)}^{D/2} \; .
\eea

Although this equation can be solved analytically for $D=3$ (as we will show in the
next section), this cannot be done in arbitrary $D$, and in general one can only obtain
parametrized solutions. As the Equation \eqref{eq:fund} is an algebraic equation it can potentially lead to
non trivial critical behaviors. We will see below that this is not the case and the partition function
and its derivatives are analytic functions in the temperature for $0\le c<1$.

The reader should keep in mind the following scenario. At fixed temperature we slowly turn on $g$.
For every $c$, there exists a critical constant $g_c(c)$ for which the tensor model reaches its
thermodynamic limit, the continuous, large volume phase. The free energy
of the continuous theory is $F(c,g_c(c))$. We will see that $F(c,g_c(c))$ is analytic in $c$ for all
$0\le c<1$, that is the continuous Ising model on a random lattice in dimension larger than three does not
undergo a phase transition at finite temperature. The numerical simulations of \cite{Ambjorn:1992iz} 
indicate that the system presents just a very long cross over between the totally ordered phase at $c=0$ and the totally disordered phase
$c=1$. We will detail the behavior of the Ising model partition function in the infinite
temperature limit $c \to 1$.

\subsubsection{The 3D case}

For $D=3$ one can cast equation \eqref{eq:fund} as
\bea
 c(U,g) =\sqrt{1- \frac{1}{U (1-gU^3)} } \Bigl[ 1-gU^3 - gU^3 \Big{(} 1- \frac{1}{U (1-gU^D)} \Big{)}  \Bigr]\crcr \; .
\eea
Introducing the variable $z = g^{1/3} U$ and squaring we obtain a cubic equation for $g^{1/3}$
\bea
c^2 =\Big{(} 1 - \frac{g^{1/3}}{z(1-z^3)}\Big{)} \Big{(} 1- 2z^3  + z^2 \frac{g^{1/3}}{ (1-z^3)} \Big{)}^2 \; ,
\eea
whose physical solution (the one which  becomes $g^{1/3} = z-z^4$ when $c = 0$) is
\bea\label{eq:gz}
 g^{1/3} = z(1-z^3) - \frac{2^2 (1-z^3)^2}{3 z^2} \sin^2 \big( \frac{\phi }{6 }\big) \; ,\qquad
 \sin\Bigl( \frac{\phi}{2} \Bigr) =  \sqrt { \frac{ 3^3 z^3  c^2  }{2^2 (1-z^3)^3 }  } \; .
\eea
It is difficult to analyze analytically the critical behavior of \eqref{eq:gz} for arbitrary $z$. In the
region of small $z$ it becomes
\bea
  g^{1/3} \approx (1-c^2) z - (1+c^2+2c^4) z^4 \;,
\eea
which, going back to the $U$ variable, leads to
\bea
 1= (1-c^2)U - (1+c^2+ 2c^4) gU^4 \rightarrow g = \frac{1-c^2}{ (1+c^2+ 2c^4) U^3} - \frac{1}{ (1+c^2+ 2c^4) U^4} \; ,
\eea
and critical behaviors for $c\to 1$
\bea
   U_c = \frac{4}{3(1-c^2 ) } \approx (1-c)^{-1}  \; ,\qquad g_c = \frac{ (1-c^2)^4 3^3}{ (1+c^2+ 2c^4) 4^4 }  \approx (1-c)^4 \; ,
\eea
hence $z_c \sim (1-c)^{1/3}$ and the initial assumption of small $z$ is self-consistent.

\subsubsection{In arbitrary dimension}

\paragraph{Change of variables.} For arbitrary dimensions one can obtain a parametrized solution of \eqref{eq:fund}. To see this we 
introduce the couple of variables
\bea
 z = g^{1/D} U \; ,\qquad w = \sqrt{1- \frac{1}{U(1-gU^D)}} \; .
\eea
As $U= \langle \bar X X \rangle$ is expected to decrease with $c$, we get $0<z\le \frac{1}{(D+1)^{1/D}}$. The physical interval of
 $w$ is readily obtained recalling \eqref{eq:VdeU}, thus $0\leq w \leq 1$.
In the $z,w$ variables we obtain the parametrized system of equations
\bea \label{maineq}
 c = (1-z^D) w - z^D w^D \ , \qquad g^{1/D} = z(1-z^D) (1-w^2) \; .
\eea
When $w=0$, one has $c=0$, and $g^{1/D} = z(1-z^D)$ which is the fundamental equation of the i.i.d. model in the
 melonic sector \cite{Bonzom:2011zz}. When $w=1$, $g=0$ and $z$ also vanishes so that $c=1$.
For $0<w<1$ we have
\bea
 \Big{(} \frac{\partial c}{\partial z} \Big{)}_{w}= - D z^{D-1} (w + w^D) <0 \; , \qquad
\Big{(}\frac{\partial c}{\partial w}\Big{)}_{z} = 1 - z^D - D z^D w^{D-1} > 0   \; .
\eea
At fixed temperature, the implicit function theorem yields
\bea
 \Big{(}\frac{\partial w}{\partial z} \Big{)}_c =  \frac{ D z^{D-1} (w+w^D) }{ 1 - z^D - D z^D w^{D-1}  } >0 \; .
\eea

\paragraph{Large volume limit and critical curve.} For sufficiently small values $g$ the free energy is analytic. We now
send the lattices to criticality. We obtain a critical curve $g_c(c)$, on which the free energy
$F(c)\equiv F(c,g_c(c))$ describes the Ising system in the large volume limit as a function of the temperature.

The critical curve is defined by the vanishing of $\partial g/\partial U$, so that the expansion of $g$ in 
term of $U$ only begins at the second order. In the $z$ parametrization it becomes 
\be
\Bigl( \frac{\partial g}{\partial z} \Bigr)_c
=  D \frac{\bigl[ z(1-z^D) (1-w^2) \bigr]^{ D-1 }}{  1 - z^D - Dz^D w^{D-1} }
\Bigl[  z^{2D} P(w) -z^D Q(w) + R(w)  \Bigr]  \; ,
\ee
with
\bea
&&  P(w) = D (1+w^2)(1+w^{D-1} ) + (1-w^2) (1 + D^2 w^{D-1})   \crcr
&&  Q(w)=D (1+w^2)(1+w^{D-1} ) +2 (1-w^2)  \crcr
&&  R(w) = 1-w^2 \; .
\eea
The roots of the first derivative write
\bea
&&  z_{1}^D (w) =  \frac{  D (1+w^2)(1+w^{D-1} ) +2 (1-w^2) - D  \sqrt{   \Big{[}  (1+w^2)^2 (1+w^{D-1} )^2 - 4   (1-w^2)^2 w^{D-1} \Big{]}  }   }
{ 2 D (1+w^2)(1+w^{D-1} ) + 2 (1-w^2) (1 + D^2 w^{D-1})  } \crcr
&&  z_{2}^D (w) =  \frac{  D (1+w^2)(1+w^{D-1} ) +2 (1-w^2) + D  \sqrt{   \Big{[}  (1+w^2)^2 (1+w^{D-1} )^2 - 4   (1-w^2)^2 w^{D-1} \Big{]}  }   }
{ 2 D (1+w^2)(1+w^{D-1} ) + 2 (1-w^2) (1 + D^2 w^{D-1})  } \crcr
&& z_3^D (w) = 0 \qquad z_4^D (w)= 1 \; .
\eea
Starting from $c=0$ and increasing the temperature, one sees that the physical solution is $z_1(w)$, being 
the only one such that $z_1^D(0) = \frac{1}{(D+1)}$.
As long as the four roots of the first derivative are well separated the Ising spins cannot undergo a second 
(or higher) order phase transition (the only way for the second derivative $\frac{\partial^2 g}{\partial z^2}$ to 
be zero is for two roots of the first derivative to collapse, as they do two dimensions where
 $z_0=z_1$ at the critical temperature \cite{Kazakov:1986hu}).
Note that for $w<1$ we have
\bea
 && 0 < D^2 \Bigl[ (1-w^2) (1- w^{\frac{D-1}{2}} )^2 + 2w^2  (1+w^{D-1} )   \Bigr]
 \Bigl[ (1-w^2) (1 +  w^{\frac{D-1}{2}} )^2 + 2w^2  (1+w^{D-1} )   \Bigr] \crcr
 && \qquad = D^2   \Big{[}  (1+w^2)^2 (1+w^{D-1} )^2 - 4   (1-w^2)^2 w^{D-1} \Big{]} \; ,
\eea
hence the first two roots can never be equal: $z_1(w) \neq z_2(w)$. Furthermore, as
 \bea
&& - D  \sqrt{   \Big{[}  (1+w^2)^2 (1+w^{D-1} )^2 - 4   (1-w^2)^2 w^{D-1} \Big{]}  }   <
 0 < D(1+w^2) - (1-w^2) \crcr
&& \qquad  < D (1+w^2)(1+w^{D-1} ) + (1-w^2) (1 + D^2 w^{D-1}) - 2 (1-w^2) \;,
\eea
we have $z_1(w) <1$. It follows that a phase transition can occur if and only if $z_1(w)$ collapses with $z_3(w)=0$.
This happens if and only if $ P(w) R(w)=0$ and, as $P(w)>D$, $z_1(w)=0   \Rightarrow w=1$ and $c=1$. 

\begin{figure}[htb]
\begin{center}
\subfigure[$(g(w),c(w))$]{ \includegraphics[scale=0.7]{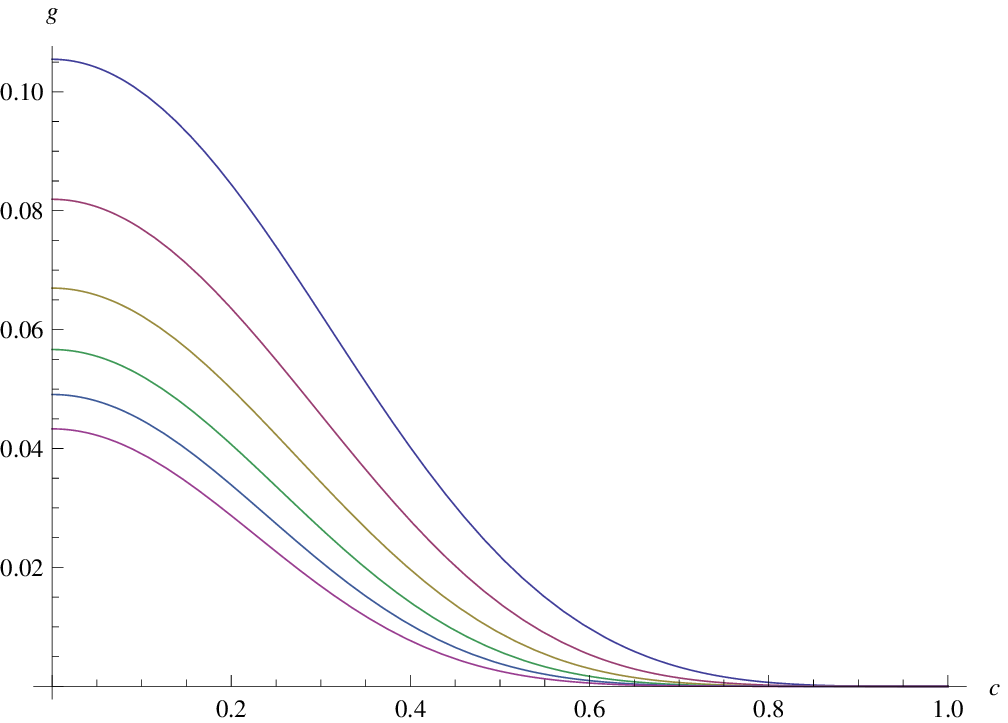}}\hspace{1cm}
\subfigure[$(g(w),U(w))$]{ \includegraphics[scale=0.7]{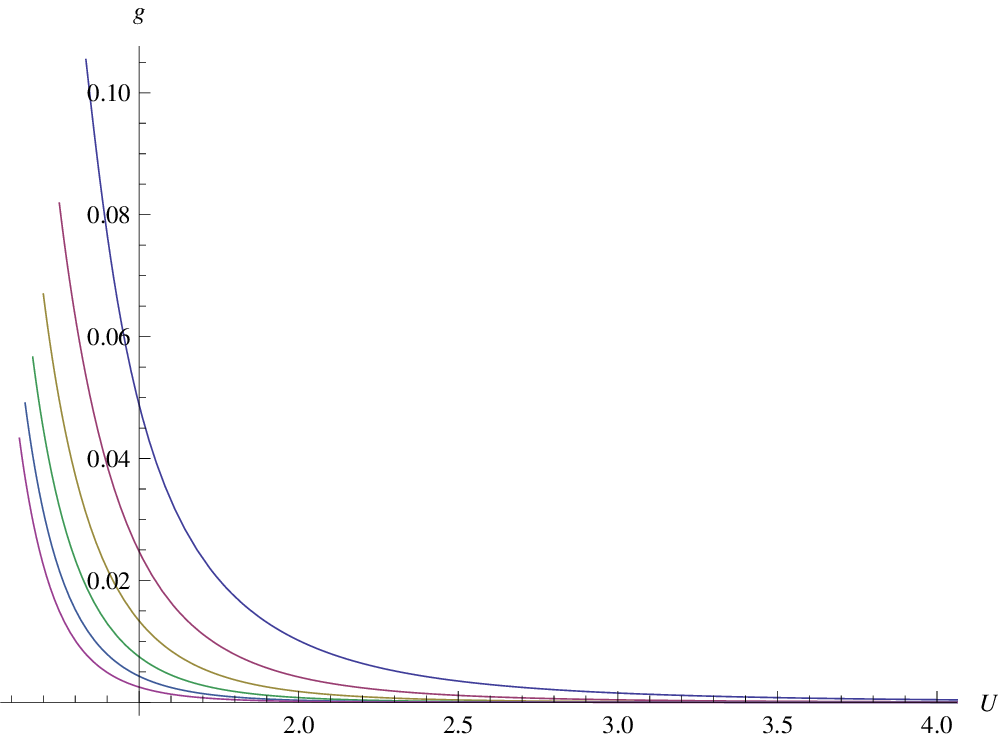}}\\
\subfigure[$(U(w),c(w))$]{ \includegraphics[scale=0.7]{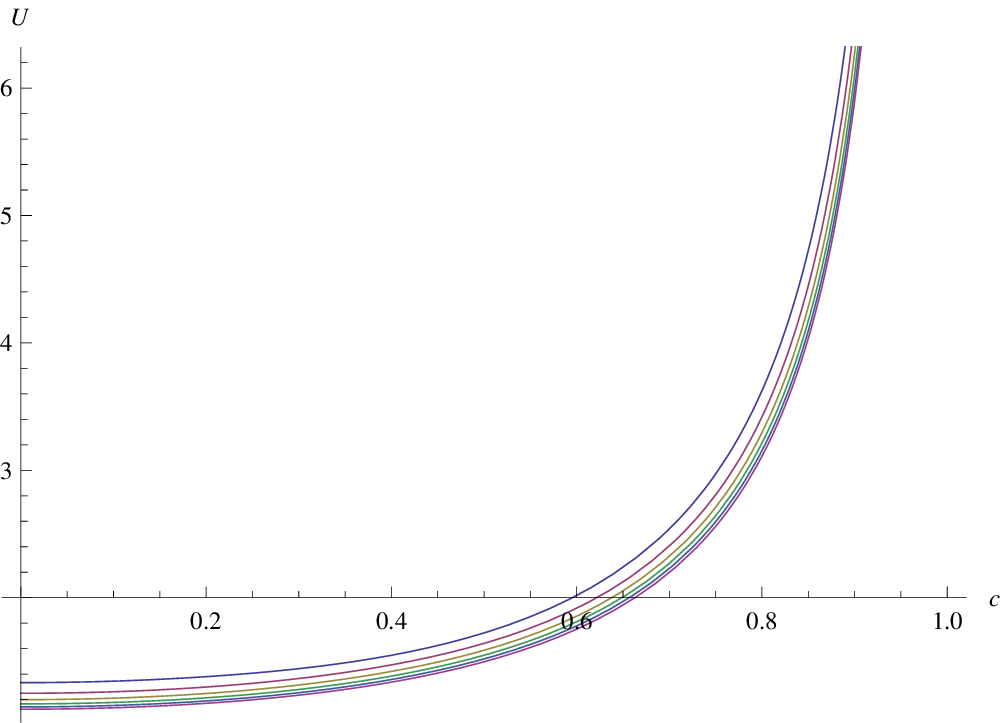}}\hspace{1cm}
\subfigure[$(V(w),c(w))$]{ \includegraphics[scale=0.7]{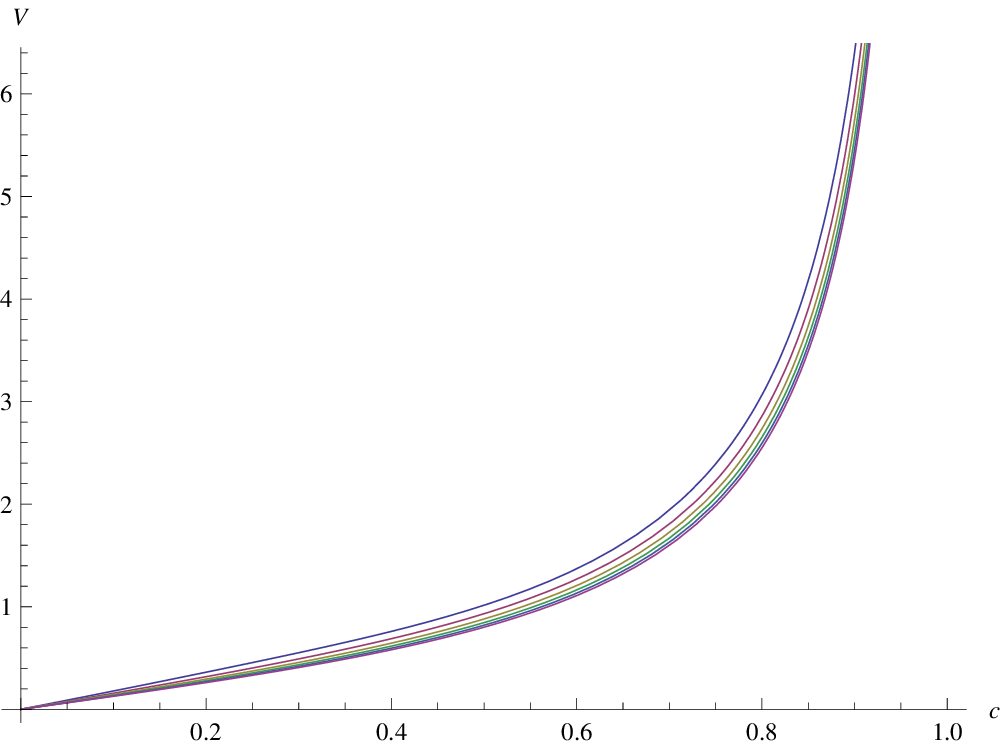}}
\caption{Parametric curves on the critical line for $D=3,\dotsc,8$.}
\label{fig:curves}
\end{center}
\end{figure}

Substituting $z_1(w)$ into \eqref{maineq} we obtain the critical curve $(c(w),g(w))$ parametrized by $w$
{\small \be
\begin{aligned}
c(w) &= \frac1{2 D w^D ( -D-1 + (D-1) w^2) -
   2 w (1 + D + ( D-1) w^2)}\Biggl[ 2(1- D^2) w^{ D+1} (1 - w^2) \\
   &\qquad{ - Dw^2 (1 + w^{D-1})(1 + w^2) \Bigl[ (1 - w^{D-1}) + \sqrt{
      (1+w^{2 D-2}) - 2 w^{D-1} \frac{1 - 6 w^2 + w^4}{(1 + w^2)^2}} \; \; \Bigr]\Biggr]}\\
g(w) &= \frac{\left(2w(1-w^2) + D(1+w^2)(w+w^D) -D \sqrt{4 w^{D+1} \left(1-w^2\right)^2+
\left(1+w^2\right)^2 \left(w+w^{D}\right)^2}\right)}{\left(2 D w^D \left((1+w^2)+D(1-w^2)\right)+
2 w \left(1+D+(-1+D) w^2\right)\right)^{D+1}\,\left(1-w^2\right)^{-D}} D^D\\
&\left((w+w^D)(1+w^2)+2Dw^D(1-w^2)
+ \sqrt{(w^2+w^{2 D}) \left(1+w^2\right)^2-2 w^D \left(w-6 w^3+w^5\right)}\right)^D \; ,
\end{aligned}
\ee}
displayed for some values of $D$ in the Figure \ref{fig:curves}. 
Using $U(w)=z_1(w)/g^{1/D}(w)$ and $V(w) = w U(w)$, one can also plot the parametric
 curves $(U(w),c(w))$, $(V(w),c(w))$ and $(g(w),U(w))$. See again the Figure \ref{fig:curves}.

\paragraph{The infinite temperature limit.} In the neighborhood of $w=c=1$, an expansion to second order gives 
$(1-c)\simeq \frac{D+1}{D}(1-w) - \frac{D+1}{2D^2}(1-w)^2$. It can be inverted to solve for the parameter $w$ 
as a function of the temperature $c$,
\be
w = 1 - \frac{D}{D+1}\,(1-c) - \frac{D}{2(D+1)^2}\,(1-c)^2 + O\bigl((1-c)^3\bigr)\;.
\ee
One can then expand all relevant quantities on the critical line around $c=1$,
\be
g_c(c)\simeq 2^{D-1}\frac{D^D}{(D+1)^{D+1}}\,(1-c)^{D+1}\;,\qquad U_c(c)\simeq V_c(c)\simeq \left(\frac{D+1}{2D}\right)\,\frac{1}{1-c}\;,
\ee
to lowest order. The derivative of the free energy along the critical curve close to $c=1$ is
\be
\begin{aligned}
\frac{dF}{dc}(c,g_c(c)) &= \frac{\partial F}{\partial c}(c,g_c(c)) + \frac{dg_c}{dc}\,\frac{\partial F}{\partial g}(c,g_c(c))\;,\\
&= -2\,V_c(c) + 2\, \frac{d\,\ln g_c}{dc} \,\bigl(1-U_c(c)+c\,V_c(c)\bigr)\;,
\end{aligned}
\ee
which gives $\frac{dF}{dc}\sim (1-c)^{-1}$.

Thus the free energy is analytic for all $0\le c <1$. In the 
{\it infinite} temperature limit $c\to 1$, the physical root $z_1(w)$ 
collapses with $z_3(w)=0$ and the free energy acquires a non analytic behavior.
The Ising spin system on random melonic lattices undergoes a phase transition at 
infinite temperature.

\section*{Acknowledgements}

Research at Perimeter Institute is supported by the Government of Canada through Industry
Canada and by the Province of Ontario through the Ministry of Research and Innovation.

\end{document}